\documentclass[12pt]{iopart}
\usepackage{epsfig,float,iopams,setstack}

\DeclareMathAlphabet{\mathsfsl}{OT1}{cmss}{m}{sl}
\begin{document}

\title[]{\bf \Large Matter collineations of Spacetime Homogeneous G\"odel-type Metrics}

\author{U. Camc{\i}$^1$ and M. Sharif$\, ^2$\footnote{Present Address:
Department of Mathematical Sciences, University of Aberdeen, Kings
College, Aberdeen AB24 3UE Scotland, UK.
$<$msharif@maths.abdn.ac.uk$>$}}
\address{$^1$ Department of Physics, Art and Science Faculty,
\d{C}anakkale Onsekiz Mart University, 17100 \d{C}anakkale,
Turkey}

\address{$^2$ Department of Mathematics, Punjab University,
Quaid-e-Azam Campus Lahore-54590, Pakistan}

\ead{ucamci@comu.edu.tr and hasharif@yahoo.com}

\begin{abstract}
The spacetime homogeneous G\"odel-type spacetimes which have four
classes of metrics are studied according to their matter
collineations. The obtained results are compared with Killing
vectors and Ricci collineations. It is found that these spacetimes
have infinite number of matter collineations in degenerate case,
i.e. det$(T_{ab}) = 0$, and do not admit proper matter
collineations in non-degenerate case, i.e. det$(T_{ab}) \ne 0$.
The degenerate case has the new constraints on the parameters $m$
and $w$ which characterize the causality features of the
G\"odel-type spacetimes.
\end{abstract}


\bigskip

\qquad \qquad \today

\section{Introduction}
\label{INT}

Let $M$ be a spacetime manifold with Lorentz metric $g$ of
signature (+ - - -). The manifold $M$ and the metric $g$ are
assumed smooth ($C^{\infty}$). Throughout this article, the usual
component notation in local charts will often be used, and a
covariant derivative with respect to the symmetric connection
$\Gamma$ associated with the metric $g$ will be denoted by a
semicolon and a partial derivative by a comma.

    Einstein's field equations (EFEs) in local coordinates are
given by
\begin{equation} \label{efe}
G_{ab} \equiv R_{ab} - \frac{1}{2} R g_{ab} = \kappa T_{ab},
\end{equation}
where $G_{ab}$ are the components of the Einstein tensor, $R_{ab}$
those of the Ricci and $T_{ab}$ of the matter (energy-momentum)
tensor. Also, $R = g^{ab} R_{ab}$ is the Ricci scalar, and it is
assumed that $\kappa = -1$ and $\Lambda = 0$ for simplicity. In
general relativity (GR) theory, the Einstein tensor $G_{ab}$ plays
a significant role, since it relates the geometry of spacetime to
its source. The GR theory, however, does not prescribe the various
forms of matter, and takes over the energy-momentum tensor
$T_{ab}$ from other branches of physics.

    The EFEs (\ref{efe}), whose fundamental constituent is the spacetime
metric $g_{ab}$, are highly nonlinear partial differential
equations, and therefore it is very difficult to obtain their
exact solutions. Symmetries of the geometrical/physical relevant
quantities of the GR theory are known as {\it collineations}. In
general, these can be represented as $\pounds_{\xi} {\cal A} =
{\cal B}$, where ${\cal A}$ and ${\cal B}$ are the
geometric/physical objects, $\xi$ is the vector field generating
the symmetry, and $\pounds_{\xi}$ signifies the Lie derivative
operator along the vector field $\xi$.

  Matter symmetries provide a different, and more physically based,
approach to symmetries of spacetimes. The physical interest of
these symmetries can be discussed as follows. For a given
distribution of matter, the contribution of gravitational
potential satisfying EFEs is the principal aim of all
investigations in gravitational physics. This has been achieved by
imposing symmetries on the geometry compatible with the dynamics
of the chosen distribution of matter. In an attempt to study the
geometric and physical properties of the electromagnetic fields,
different types of collineations have been investigated
\cite{ahsan1},\cite{ahsan2} along with many other interesting
results. It is seen that for a null electromagnetic field a motion
does not imply Maxwell collineation and conversely. Symmetries of
the energy-momentum tensor (also called matter collineations
defined in the following) provide conservation laws on matter
fields. These enable us to know how the physical fields, occupying
in certain region of spacetimes, reflect the symmetries of the
metric \cite{coley}. In other words, given the metric tensor of a
spacetime, one can find symmetry for the physical fields
describing the material content of that spacetime.

    A one-parameter group of conformal motions generated by
a {\it conformal Killing vector} (CKV) $\xi$ is defined as
\cite{katzin}
\begin{equation}
\pounds_{\xi} g_{ab} = 2 \psi g_{ab}, \label{confm}
\end{equation}
where $\psi=\psi(x^a)$ is a conformal factor. If $\psi_{;\, a b}
\neq 0$, the CKV is said to be {\it proper}. Otherwise, $\xi$
reduces to the special {\it conformal Killing vector} (SCKV) if
$\psi_{;\,ab} = 0$, but $\psi_{,a}\neq 0$. Other subcases are {\it
homothetic vector} (HV) if $\psi_{,a}=0$ and {\it Killing vector}
(KV) if $\psi= 0$.

Using Eq. (\ref{confm}), we find from Eq. (\ref{efe}) that
\begin{equation} \label{lietab}
\pounds_{\xi} T_{ab} = - 2 \psi_{;ab} + 2 g_{ab} \Box \psi  ,
\end{equation}
where $\Box$ is the Laplacian operator defined by $\Box \psi
\equiv g^{cd} \psi_{;cd}$. Therefore, for a KV, or HV, or SCKV we
have
\begin{equation} \label{mcol1}
\pounds_{\xi} T_{ab} = 0 \quad \Leftrightarrow \quad \pounds_{\xi}
G_{ab} = 0,
\end{equation}
or in component form
\begin{equation} \label{mcol2}
\quad T_{ab,c} \xi^c + T_{ac} \xi^c_{,b} + T_{cb} \xi^c_{,a} = 0.
\end{equation}
A vector field $\xi$ satisfying Eq. (\ref{mcol1}) or (\ref{mcol2})
on $M$ is called a {\it matter collineation} (MC). Therefore, we
define a {\it proper} MC to be an MC which is not an KV, {\it or}
an HV, {\it or} an SCKV. Since the Ricci tensor arises naturally
from the Riemann curvature tensor (with components $R^a_{\, bcd}$
and where $R_{ab} \equiv R^c_{\, acb}$) and hence from the
connection, the study of {\it Ricci collineation} (RC) defined by
$\pounds_{\xi} R_{ab} = 0$ has a natural geometrical significance
\cite{collinson1}-\cite{cnp}. Mathematical similarities between
the Ricci and energy-momentum tensors mean many techniques for
their study should show some similarities. From the physical
viewpoint explained in the above, a study of MCs, i.e. to look
into the set of solutions to Eq.(\ref{mcol2}) seems more relevant.
In this direction, some papers have recently been appeared on MCs
\cite{ccv}-\cite{seher}. In addition, since energy-momentum tensor
$T_{ab}$ is more fundamental in the study of dynamics of fluid
spacetimes of GR, the remainder of this paper will be concerned
with MCs.

Recently, Carot {\it et al.} \cite{ccv} and Hall {\it et al.}
\cite{hall} have noticed some important results about the Lie
algebra of MCs. These are the following:

\begin{description}
\item[a.]  \quad The set of all MCs on $M$ is a vector space, but it may be
infinite dimensional and may not be a Lie algebra. If $T_{ab}$ is
non-degenerate, i.e. $det(T_{ab}) \neq 0$, the Lie algebra of MCs
is finite dimensional. If $T_{ab}$ is degenerate, i.e.
$det(T_{ab}) = 0$, we cannot guarantee the finite dimensionality
of the MCs.

\item[b.] \quad If the energy-momentum tensor $T_{ab}$ is everywhere
of rank $4$ then it may be regarded as a metric on $M$. Then, it
follows by a standard result that the family of MCs is, in fact, a
Lie algebra of smooth vector fields on $M$ of finite dimension
$\leq 10$ (and $\neq 9$).

\item[c.] \quad If a vector field $\xi$ on $M$ is a symmetry of {\it all} the
gravitational field sources, then one could require the Eq.
(\ref{mcol1}) (for the {\it non-vacuum} sources) and
$\pounds_{\xi} C^a_{bcd} = 0$ (for the {\it vacuum} sources),
where $C^a_{bcd}$ are Weyl curvature tensor components.
\end{description}

Throughout this paper, it is used the form (\ref{mcol2}) of MCs
without imposing any restriction on the energy-momentum tensor.

    The plan of the paper is as follows. In the next section we shall
describe the G\"odel-type spacetimes with some general results. In
Section \ref{MC}, we shall write down MC Eqs. (\ref{mcol2}) for
the metric given by (\ref{metric}), and solve them for spacetime
homogeneous G\"odel-type metrics. Finally, we shall provide a
brief summary and discussion of the results obtained.

\section{G\"odel-type metrics}
\label{ST}

In 1949, G\"odel found a solution of Einstein's field equations
with cosmological constant for incoherent matter with rotation
\cite{godel}. It is certainly the best known example of a
cosmological model which makes it apparent that GR does not
exclude the existence of closed timelike world lines, despite its
Lorentzian character which leads to the local validity of the
causality principle. G\"odel's cosmological solution has a
well-recognized importance which has, to a large extent, motivated
the investigations on rotating cosmological G\"odel-type
spacetimes and on causal anomalies within the framework of GR
\cite{som,banerjee}. In natural cylindrical coordinates $x^a =
(t,r,\phi,z), \, a= 0,1,2,3$, the G\"odel-type metrics are given
by
\begin{equation}\label{metric}
ds^2 =  \left[ dt + H(r) d\phi \right]^2 - dr^2 - D^2(r) d\phi^2 -
dz^2.
\end{equation}
In 1980, Raychaudhuri and Thakurta \cite{rayc} has given the
necessary conditions that a metric of G\"odel-type is spacetime
homogeneous (ST homogeneous, hereafter). Three years later,
Rebou\c{c}as and Tiomno \cite{rebo1} proved that these conditions
\begin{eqnarray}
& & \frac{D''}{D} = const \equiv m^2 , \label{cond1} \\& &
\frac{H'}{D} = const \equiv - 2\omega \label{cond2}
\end{eqnarray}
are both necessary and sufficient. However, in both articles
\cite{rayc,rebo1}, the study of ST homogenity is limited in that
only time-independent  KV fields were considered \cite{teix}.
Finally, these conditions were proved to be the necessary and
sufficient conditions for a G\"odel-type manifold to be ST
homogeneous without assuming any such simplifying hypothesis
\cite{rebo2}. The above results for G\"odel-type manifolds can be
collected together as follows :

{\bf Theorem 1}: The necessary and sufficient conditions for a
four-dimensional Riemannian G\"odel-type manifold to be locally
homogeneous are those given by Eqs. (\ref{cond1}) and
(\ref{cond2}).

{\bf Theorem 2}: The four-dimensional homogeneous Riemannian
G\"odel-type manifolds admit group of isometry $G_r$ with

(i) $r = 5$ if $m^2 (>0) \neq 4 \omega^2$ with $\omega \neq 0$, or
when $m^2 = 0$ and $\omega \neq 0$, or when $m^2 \equiv - \mu^2 (<
0)$ and $\omega \neq 0$;

(ii) $r = 6$ if $m^2 \neq 0$ and $\omega = 0$;

(iii) $r = 7$ if $m^2 (>0) = 4 \omega^2$ and $\omega \neq 0$.

{\bf Theorem 3}: The four-dimensional homogeneous Riemannian
G\"odel-type manifolds are locally characterized by two
independent parameters $m^2$ and $\omega$: the pair of ($m^2,
\omega$) identically specify locally equivalent manifolds.

Now, we shall be concerned with irreducible set of isometrically
nonequivalent homogeneous G\"odel-type metrics which was given in
Refs. $28$ and  $29$. These distinguish the following four classes
of metrics according to:

{\it Class I} : $m^2 > 0, \omega  \neq 0$. For this case, the
general solution of Eqs. (\ref{cond1}) and (\ref{cond2}) can be
written as
\begin{equation}
H(r) = \frac{2 \omega}{m^2} \left[ 1 - \cosh(mr) \right] \quad and
\quad D(r) = \frac{1}{m} \sinh(mr). \label{class1}
\end{equation}

{\it Class II} : $m^2 = 0, \omega \neq 0$. For this case, the
general solution of Eqs. (\ref{cond1}) and (\ref{cond2}) is
\begin{equation}
H(r) = - \omega r^2 \quad and \quad D(r) = r, \label{class2}
\end{equation}
where only the essential parameter $\omega$ appears.

{\it Class III} : $m^2 \equiv - \mu^2 < 0, \omega \neq 0$.
Similarly, the integration of the conditions for homogeneity Eqs.
(\ref{cond1}) and (\ref{cond2}) leads to
\begin{equation} H(r) =
\frac{2 \omega}{\mu^2} \left[ \cos(\mu r) -1 \right] \quad and
\quad D(r) = \frac{1}{\mu} \sin(\mu r). \label{class3}
\end{equation}

{\it Class IV} : $m^2 \neq 0, \omega = 0$. We refer to the
manifolds of this class as degenerated G\"odel-type manifolds,
since the cross term in the line element, related to the rotation
$\omega$ in the G\"odel model, vanishes. By a trivial coordinate
transformation one can make $H = 0$ with $D(r)$ given,
respectively, by Eqs. (\ref{class1}) or (\ref{class3}) depending
on whether $m^2 > 0$ or $ m^2 \equiv - \mu^2 < 0$.

If $m^2 = \omega = 0$, then the line element (\ref{metric}) is
clearly Minkowskian. Therefore, this particular case has not been
included in this study. Also, it is noted that the condition $m^2
= 2 w^2$ defines nothing but the original G\"odel model, which is
known to violate the causality principle. Rebou\c{c}as {\it et
al.} \cite{rebo2} and Calv\~{a}o {\it et al.} \cite{calvao} have
found that the causality features of the G\"{o}del-type spacetimes
depend upon the the two independent parameters given above, i.e.
$m$ and $w$. They have shown that for $0 \leq m^2 < 4 w^2$, there
exists only \emph{one} noncausal region; for $m^2 \geq 4 w^2$
there are no closed timelike curves, in which a completely causal
and ST homogeneous G\"odel model corresponds to the limiting case
$m^2 = 4 w^2$; for $m^2 < 0$ there are infinite number of
alternating causal and noncausal regions.

The KV fields as well as corresponding Lie algebra of each class
are given in Appendix B \cite{rebo1},\cite{teix},\cite{rebo4}.
Firstly, Hall and Costa \cite{hall-costa} have pointed out that
the original G\"odel metric does not admit HVs and laterly, this
result is extended to the ST homogeneous G\"odel-type spacetimes
\cite{melfo}. Recently, the proper CKVs and complete conformal
algebra of a G\"odel-type spacetime have been computed in
\cite{tsamparlis2} applying their method. The RCs and contracted
RCs of ST homogeneous G\"odel-type spacetimes are studied by Melfo
{\it et al.} \cite{melfo}. In this study, we provide a complete
solution of the MC equations for ST homogeneous G\"odel-type
spacetimes.

\section{Matter Collineation Equations and Their Solutions}
\label{MC}

For the ST homogeneous G\"odel-type metrics, the non-vanishing
components of $T_{ab}$ become
\begin{eqnarray}
& & T_{00} = 3 w^2 - m^2, \label{t00} \\& & T_{02} = (3w^2 - m^2)
H, \label{t02} \\& & T_{11} = w^2, \label{t11} \\& & T_{22} =
(3w^2 - m^2) H^2 + w^2 D^2, \label{t22}
\\& & T_{33} = m^2 - w^2, \label{t33}
\end{eqnarray}
(see Appendix A). Thus, $det(T_{ab}) = w^4 (m^2 - w^2)(3w^2 - m^2)
D^2$. If $w^2 \neq 0,\, m^2 \neq m^2$,  {\it and} $m^2 \neq 3
w^2$, then $T_{ab}$ is non-degenerate. When $w^2 = 0$ {\it or}
$m^2 = w^2$ {\it or} $m^2 = 3 w^2$, then $T_{ab}$ is degenerate.

For the ST homogeneous G\"odel-type metric (\ref{metric}), writing
equation (\ref{mcol2}) in expanded form, we obtain the following
MC equations :
\begin{eqnarray}
& & w^2 \xi^1_{,r} = 0, \label{mc-a}
\\& & (m^2 - w^2) \xi^3_{,z} = 0, \label{mc-b} \\& & w^2 \xi^1_{,z}
+ (m^2 -w^2)\xi^3_{,r} = 0, \label{mc-c} \\& & (3 w^2 -m^2) F_{,t}
= 0, \label{mc-d} \\& & (3w^2 -m^2) F_{,z} +(m^2 - w^2)\xi^3_{,t}
= 0, \label{mc-e} \\ & & (3 w^2 - m^2) H\, F_{,z} + w^2 D^2
\xi^2_{,z} + (m^2 - w^2) \xi^3_{,\phi} = 0, \label{mc-f}
\\& & (3 w^2 - m^2) H \left[F_{,\phi} + H' \xi^1 \right]+ w^2 D^2 \left[ \xi^2_{,\phi} +
\frac{D'}{D} \xi^1 \right] = 0, \label{mc-g}
\\& & (3w^2 - m^2) \left[ F_{,\phi} - 2 w D \xi^1 \right] + w^2
D^2 \xi^2_{,t} = 0, \label{mc-h}
\\& & (3 w^2 - m^2) \left[ F_{,r} + 2w D \xi^2 \right] + w^2
\xi^1_{,t} = 0, \label{mc-i}
\\& & (3 w^2 - m^2) H\, \left[ F_{,r} + 2w D \xi^2 \right]
+ w^2 D^2 \xi^2_{,r} + w^2 \xi^1_{,\phi} = 0. \label{mc-j}
\end{eqnarray}
where $F$ is defined as
\begin{equation}
F = \xi^0 + H \xi^2.
\end{equation}
Now, in the following, we consider degenerate and non-degenerate
cases of MCs, respectively.

\subsection{Degenerate Case}
\label{degenerate}

In this case, we have the possibilities in which the
energy-momentum tensor $T_{ab}$ becomes degenerate, ({\bf i}) $w
=0, m^2 \neq 0$; ({\bf ii}) $m^2 = w^2, w \neq 0$; ({\bf iii})
$m^2 = 3 w^2, w \neq 0$. It is seen from these possibilities that
we do {\it not} have degenerate energy-momentum tensor $T_{ab}$ in
the Class II.

{\bf Case (i)}. This case corresponds to the Class IV case, and
since one can make $H = 0$ by a trivial coordinate transformation,
we obtain
\begin{equation}
T_{00} = -m^2 = - T_{33}, \qquad T_{02} = T_{11} = T_{22} = 0.
\end{equation}
The solution of the MC equations gives
\begin{equation}
\xi^0 = c_1 + c_3 z, \,\, \xi^1 = \xi^1 (t,r,\phi,z), \,\, \xi^2 =
\xi^2 (t,r,\phi,z), \,\, \xi^3 = c_2 + c_3 t,
\end{equation}
which can be written as follows
\begin{eqnarray}
& & {\bf \xi}_{(1)} = \partial_t, \quad {\bf \xi}_{(2)} =
\partial_{z}, \quad {\bf \xi}_{(3)} = z \partial_t + t \partial_z,  \label{deg-a} \\& & {\bf
\xi}_{(4+i)} = \xi^1 (t,r,\phi,z) \partial_r + \xi^2 (t,r,\phi,z)
\partial_{\phi}, \label{deg-b}
\end{eqnarray}
where ``{\it i}'' is an arbitrary positive number. In order to
construct a closed abelian algebra for vectors given in
(\ref{deg-a}) and (\ref{deg-b}), we find that $\xi^1 = \xi^1
(r,\phi)$ and $\xi^2 = \xi^2 (r,\phi)$. Thus, we have infinite
number of MCs.

{\bf Case (ii)}. For this case, the Classes I and III are to be
taken into consideration. The condition $m^2 = w^2$ give rise to
\begin{equation}
T_{00} = 2 w^2, \,\, T_{02} = 2 w^2 H, \,\, T_{11} = w^2, \,\,
T_{22} = w^2 ( 2 H^2 + D^2), \,\, T_{33} = 0.
\end{equation}
Then, the MC equations yield the solutions
\begin{eqnarray}
\xi^0 = c_1 - c_2 \left( \frac{2}{w} D' + H \right), \,\, \xi^1 =
0, \,\, \xi^2 = c_2, \,\, \xi^3 = \xi^3 (t,r,\phi,z),
\end{eqnarray}
which can be stated as
\begin{eqnarray}
& & {\bf \xi}_{(1)} = \partial_t , \quad {\bf \xi}_{(2)} =
\partial_{\phi} - \left( \frac{2}{w} D' + H \right) \partial_t, \quad {\bf \xi}_{(3 +
i)}= \xi^3 (t,r,\phi,z) \partial_z .
\end{eqnarray}
Therefore, we have also infinite number of MCs.

{\bf Case (iii)}. In this case,  the MC equations reveal only for
Classes I and III that
\begin{eqnarray}
& & \xi^0 = \xi^0 (t,r,\phi,z), \quad \xi^1 = c_1 \cos\phi + c_2
\sin\phi,  \\& & \xi^2 = \frac{D'}{D}\left( -c_1 \sin\phi + c_2
\cos\phi \right) + c_3, \quad \xi^3 = c_4.
\end{eqnarray}
Then, it follows from these results that
\begin{eqnarray}
& & {\bf \xi}_{(1)} = \xi^0 (t,r,\phi,z) \partial_t, \\& &  {\bf
\xi}_{(2)} = \partial_z + \xi^0 (t,r,\phi,z) \partial_t, \\& &
{\bf \xi}_{(3)} = \partial_{\phi} + \xi^0 (t,r,\phi,z)
\partial_t, \\& & {\bf \xi}_{(4)} = \xi^0 (t,r,\phi,z) \partial_t + \cos\phi \partial_r - \frac{D'}{D}\sin\phi
\partial_{\phi}, \\& & {\bf \xi}_{(5)} = \xi^0 (t,r,\phi,z) \partial_t - \sin\phi \partial_r - \frac{D'}{D}\cos\phi
\partial_{\phi},
\end{eqnarray}
where we have used the following property
\begin{equation}
D^2 \left(\frac{D'}{D}\right)' = -1,
\end{equation}
which is valid for ST homogeneous G\"odel-type metrics only. Thus,
in this case one also finds infinite number of MCs. For this case,
the constraint $m^2 = 3 w^2$ yields
\begin{equation}
T_{11} = w^2, \quad T_{22} = w^2 D^2, \quad T_{33} = 2 w^2 \qquad
T_{00} = T_{02} = 0.
\end{equation}

\subsection{Non-degenerate Case}
\label{nondegenerate}

In this case, since the energy-momentum tensor $T_{ab}$ is
non-degenerate, we assume that $w^2 \neq 0, m^2 \neq w^2$ and $m^2
\neq 3 w^2$. From Eqs. (\ref{mc-a})-(\ref{mc-g}), it follows that
\begin{eqnarray}
\xi^0&= &-H \, \xi^2 + \frac{z}{3w^2 - m^2} (c_1 \phi +
c_2) + P(r,\phi), \nonumber \\ \xi^1 &=& Q(t,\phi), \nonumber \\
\xi^2 &=& \frac{z}{w^2 D^2} \left[ c_1 (t-\phi
\,H) - c_2 \, H - c_3 (m^2 - w^2) \right] - \frac{3W^2 -m^2}{w^2})\frac{H}{D^2} P(r,\phi)  \nonumber \\
\qquad & & + \left[ \frac{2}{w}(3w^2 -m^2) \frac{H}{D} -
\frac{D'}{D} \right] \int{Q(t,\phi) d\phi} + R(t,r), \nonumber
\\ \xi^3 & = & \frac{t}{w^2 - m^2}(c_1 \phi + c_2) t + c_3 \phi + c_4, \nonumber
\end{eqnarray}
where $P(r,\phi), Q(t,\phi)$, and $R(t,r)$ are integration
functions. Then, using Eqs. (\ref{mc-h}) and (\ref{mc-i}), we find
for Classes I, II, and III that $c_1 = c_2 = c_3 = 0$ and
\begin{eqnarray}
& & P = 2 w D \int{K_1 (\phi) d^2 \phi} + 2 w \phi \, D \, c_5 +
L_1 (r), \\& & Q = \int{K_1(\phi) d\phi} + c_5, \quad R = L_2 (r),
\\& & L_{1,r} - \frac{2}{w} (3w^2 - m^2) \frac{H}{D} L_1 + 2w D
L_2 = 0, \label{ceq1}
\end{eqnarray}
where $K_1 (\phi), L_1 (r)$, and $L_2 (r)$ are integration
functions, and we have assumed $m^2 \neq 4 w^2$. Then, from the
remaining Eq. (\ref{mc-j}), we obtain that $c_5 = 0$ and
\begin{eqnarray}
& & K_{1,\phi \phi} - \epsilon K_1 = 0, \label{ceq2}
\\& & L_{2,r} - (3 - \frac{m^2}{w^2})\left( \frac{H}{D^2} L_1
\right)_{,r} = 0, \label{ceq3} \\& & D^2 \left( \frac{D'}{D}
\right)' = \epsilon. \label{ceq4}
\end{eqnarray}
For the ST homogeneous G\"odel-type metrics in cylindrically
coordinates, $\epsilon$ takes the value $-1$ only, while $\epsilon
= 0$ case coincides with the original G\"odel metric \cite{godel}.
The solution of the above equations (\ref{ceq1})-(\ref{ceq4}) for
Class I is given as follows
\begin{eqnarray}
& & \xi^0 = \frac{H}{D} (k_1 \cos\phi + k_2 \sin\phi) -
\frac{2w}{m}k_3 + k_4,  \nonumber \\& & \xi^1 = k_1 \sin\phi - k_2
\cos\phi, \nonumber \\& & \xi^2 = \frac{D'}{D} (k_1 \cos\phi + k_2
\sin\phi) + m k_3, \nonumber
\\& & \xi^3 = k_5, \nonumber
\end{eqnarray}
where $k_1, k_2, k_3, k_4,$ and $k_5$ are constants. Thus, these
correspond to the KVs for this class (see Appendix B). Similarly,
using Eqs.(\ref{ceq1})-(\ref{ceq4}), one can obtain the MCs for
Classes II and III which are only the KVs.

When $m^2 = 4 w^2$, Eqs. (\ref{mc-i}) and (\ref{mc-j}) yield that
$c_1 = c_2 = c_3 = 0$, and
\begin{eqnarray}
& & P = 2 w D \int{K_1 (\phi) d^2 \phi} + L_1 (r), \quad R = L_2
(r), \\& & Q_{,tt} + m^2 Q = m^2 \int{K_1 d\phi}, \\& & Q_{,\phi
\phi} + Q = K_{1,\phi} + \int{K_1 d\phi},
\\& & K_{1,\phi \phi} + K_1 = 0,  \\& & L_{1,r} + 2 w \frac{H}{D}
L_1 + 2 w D L_2 = 0, \\& & \frac{H}{D^2} L_1 + L_2 = c_5.
\end{eqnarray}
Now, after solving the above equations and rearranging the
appearing constants, it follows that
\begin{eqnarray}
& & \xi^0 = \frac{H}{D} \left[ k_1 \sin(m t + \phi) - k_2 \cos(m t
+ \phi) + k_3 \cos\phi + \sin\phi \right] - k_6 + k_7, \nonumber
\\& & \xi^1 = k_1 \cos(m t + \phi) + k_2 \sin(m t + \phi) + k_3
\sin\phi - k_4 \cos\phi, \nonumber \\ & & \xi^2 = - \frac{1}{D}
\left[ k_1 \sin(m t + \phi) - k_2 \cos(m t + \phi) \right]
\nonumber \\ & & \qquad
\quad + \frac{D'}{D}(k_3 \cos\phi + k_4 \sin\phi) + m k_6, \nonumber  \\
& & \xi^3 = k_5 \nonumber
\end{eqnarray}
which are the KVs given in Class I for case $m^2 = 4 w^2$.

\section{Conclusions}
\label{CONC}

In this paper, we have obtained MCs for ST homogeneous
G\"odel-type metrics. As a conclusion, the RC and MC equations for
these metrics deserve some remarks. First, for these spacetimes,
we have only degenerate case of the RC equations, because of
det$(R_{ab}) = 0$. Also, $\xi^3$ does not appear in the RC
equations, and therefore any metric of the form (\ref{metric})
admits a proper RC of the form \cite{melfo} ${\bf \xi} =
f(t,r,\phi,z) \partial_z$ which represents that we have infinite
number of RCs. Second, for the ST homogeneous G\"odel-type
metrics, we have both degenerate and non-degenerate cases of the
MC equations given in Sec.3. In the MC classification of these
spacetimes according to the nature of the energy-momentum tensor,
we find that when the energy-momentum tensor is degenerate,
subsection 3.1, then there are infinite number of MCs. In this
subsection, we have found the new constraints on the parameters
$m$ and $w$, which are $m^2 = w^2$, case (ii), and $m^2 = 3 w^2$,
case (iii). These new constraints are in the interval $0 < m^2 <
4w^2$, which means that there exist only one noncausal region.
Also, the cases (ii) and (iii) correspond to the Classes I and III
only. For the case (ii), we have from Eq. (\ref{rskaler}) that the
Ricci scalar vanishes. Therefore, the MCs and RCs are obviously
coincide for this case. Furthermore, in subsection 3.2, we proceed
to deal with the MCs of the ST homogeneous G\"odel-type spacetimes
in which the energy-momentum tensor is non-degenerate. In this
subsection, we find that all MCs of ST homogeneous G\"odel-type
metrics are the KVs given in Appendix B, that is, these spacetimes
given by the Classes I-IV do not admit proper MCs.

Although it is noted in Ref. $13$ that when rank of $T_{ab}$ is
{\it one}, the physical interest in degenerate case is only
limited to dust fluids (perfect fluids with $p=0$), {\it or} null
fluids (radiation and null Einstein-Maxwell fields), it seems that
when rank of $T_{ab}$ is {\it three}, i.e. the cases (ii) and
(iii), it may be possible to find the physically significant
spacetimes with degenerate energy-momentum tensor.

\section*{Acknowledgements}

We would like to thank for the travelling grant of ICTP Net-53
under the BIPTUN programme, Center for Physics, Quad-i Azam
University, Islamabad-Pakistan and \d{C}anakkale Onsekiz Mart
University, \d{C}anakkale-Turkey, for the local hospitality.

\newpage

\renewcommand{\theequation}{A\arabic{equation}}
\setcounter{equation}{0}
\section*{Appendix A}

For the G\"odel-type space-times, we find the following results
for the non-vanishing components of the fully-covariant curvature
tensor $R_{abcd}$
\begin{eqnarray}
& & R_{0101} = \left( \frac{H'}{2D} \right)^2, \qquad R_{0202} =
\frac{{H'}^2}{4},  \label{rie1} \\& & R_{0112} = - \frac{H}{2}
\left[ \frac{D}{H} \left( \frac{H'}{D} \right)' + \frac{1}{2}
\left( \frac{H'}{D} \right)^2 \right], \label{rie2} \\& & R_{1212}
= H^2 \left[ \frac{D}{H} \left( \frac{H'}{D} \right)' + \left(
\frac{H'}{2D} \right)^2 + 3 \left( \frac{H'}{2 H} \right)^2
\right] - D D'', \label{rie3}
\end{eqnarray}
the Ricci tensor $R_{ab}$
\begin{eqnarray}
& & R_{00} = \frac{1}{2} \left( \frac{H'}{D} \right)^2, \qquad
R_{11} = \frac{D''}{D} - \frac{1}{2} \left( \frac{H'}{D}
\right)^2,  \label{ricci00-11} \\& & R_{02} =  - \frac{H}{2}
\left[ \frac{D}{H} \left( \frac{H'}{D} \right)' + \left(
\frac{H'}{D} \right)^2 \right], \label{ricci02} \\& & R_{22} = -
H^2 \left[ \frac{D}{H} \left( \frac{H'}{D} \right)' + \frac{1}{2}
\left( \frac{H'}{D} \right)^2 + \frac{1}{2} \left( \frac{H'}{H}
\right)^2 \right] + D D'', \label{ricci22}
\end{eqnarray}
and the scalar curvature $R$
\begin{equation} R = - \frac{2D''}{D}
+ \frac{1}{2} \left( \frac{H'}{D} \right)^2 \label{rskaler}
\end{equation}
respectively. By using the above expressions, the non-vanishing
components of the Einstein tensor $G_{ab}$ are the following
\begin{eqnarray}
& & G_{00} = \frac{D''}{D} - 3 \left( \frac{H'}{2D} \right)^2 ,
\qquad G_{11} = - \left( \frac{H'}{2D} \right)^2, \label{ein00-11}
\\& & G_{02} = - \frac{H}{2} \left[ \frac{D}{H} \left(
\frac{H'}{D} \right)' - 2 \frac{D''}{D} + \frac{3}{2} \left(
\frac{H'}{D} \right)^2 \right], \label{ein02} \\& & G_{22} = - H^2
\left[ \frac{D}{H} \left( \frac{H'}{D} \right)' - \frac{D''}{D} +
3 \left( \frac{H'}{2 D} \right)^2 + \left( \frac{H'}{2 H}
\right)^2 \right], \label{ein22} \\& & G_{33} = - \frac{D''}{D} +
\left( \frac{H'}{2D} \right)^2.
\end{eqnarray}

\renewcommand{\theequation}{B\arabic{equation}}
\setcounter{equation}{0}
\section*{Appendix B}

The KVs of the ST homogeneous G\"odel-type spacetimes presented in
the following were first obtained in Refs. $24$ and $25$, can also
be derived as a particular case of KVs calculated in Ref. $29$.
Thus, Linearly independent KVs for the G\"odel-type space-times in
Class I-Class IV are given as

{\it Class I} : $m^2 > 0, \omega  \neq 0$. When $m^2 \neq 4
\omega^2$,
\begin{eqnarray}
& & {\bf K}_1 = \partial_t, \quad {\bf K}_2 = \partial_z, \quad
{\bf K}_3 = \frac{2 \omega}{m} \partial_t - m \partial_{\phi},
\nonumber
\\ & & {\bf K}_4 = - \frac{H}{D} \sin\phi \partial_t + \cos\phi
\partial_r - \frac{D'}{D} \sin\phi \partial_{\phi}, \\
& & {\bf K}_5 = - \frac{H}{D} \cos\phi \partial_t - \sin\phi
\partial_r - \frac{D'}{D} \cos\phi \partial_{\phi} \nonumber
\end{eqnarray}
The Lie algebra has the following nonvanishing commutators:
\begin{eqnarray}
& & \left[ {\bf K}_3, {\bf K}_4 \right] = - m {\bf K}_5, \quad
\left[ {\bf K}_4, {\bf K}_5 \right] = m {\bf K}_3, \quad \left[
{\bf K}_5, {\bf K}_3 \right] = - m {\bf K}_4 . \nonumber
\end{eqnarray}
It should be noticed that the expressions for all KVs are
time-independent. When $m^2 = 4 \omega^2$,
\begin{eqnarray}
& & {\bf K}_1 = \partial_t, \quad {\bf K}_2 = \partial_z, \quad
{\bf K}_3 = \partial_t - m \partial_{\phi}, \nonumber
\\ & & {\bf K}_4 = - \frac{H}{D} \sin\phi \partial_t + \cos\phi
\partial_r - \frac{D'}{D} \sin\phi \partial_{\phi},  \nonumber \\
& & {\bf K}_5 = - \frac{H}{D} \cos\phi \partial_t - \sin\phi
\partial_r - \frac{D'}{D} \cos\phi \partial_{\phi}, \\&
& {\bf K}_6 = -\frac{H}{D} \cos(m t + \phi) \partial_t + \sin(m t
+ \phi) \partial_r + \frac{1}{D} \cos(m t + \phi) \partial_{\phi},
\nonumber \\& & {\bf K}_7 = -\frac{H}{D} \sin(m t + \phi)
\partial_t - \cos(m t + \phi) \partial_r + \frac{1}{D} \sin(m t +
\phi) \partial_{\phi}, \nonumber
\end{eqnarray}
where $m = + 2\omega$. The corresponding Lie algebra has the
following nonvanishing commutators:
\begin{eqnarray}
& & \left[ {\bf K}_3, {\bf K}_4 \right] = - m {\bf K}_5, \quad
\left[ {\bf K}_4, {\bf K}_5 \right] = m {\bf K}_3, \quad \left[
{\bf K}_5, {\bf K}_3 \right] = - m {\bf K}_4 , \nonumber \\& &
\left[ {\bf K}_1, {\bf K}_6 \right] = - m {\bf K}_7, \quad \left[
{\bf K}_6, {\bf K}_7 \right] = m {\bf K}_1, \quad \left[ {\bf
K}_7, {\bf K}_1 \right] = - m {\bf K}_6 . \nonumber
\end{eqnarray}

{\it Class II} : $m^2 = 0, \omega \neq 0$. In this Class, the KVs
are
\begin{eqnarray}
& & {\bf K}_1 = \partial_t, \quad {\bf K}_2 = \partial_z, \quad
{\bf K}_3 = \partial_{\phi}, \nonumber
\\ & & {\bf K}_4 = - \omega \, r \, \sin\phi \partial_t - \cos\phi
\partial_r + \frac{1}{r} \sin\phi \partial_{\phi}, \\
& & {\bf K}_5 = - \omega \, r \, \cos\phi \partial_t + \sin\phi
\partial_r + \frac{1}{r} \cos\phi \partial_{\phi} \nonumber
\end{eqnarray}
For these KVs, the Lie algebra is
\begin{eqnarray}
& & \left[ {\bf K}_3, {\bf K}_4 \right] = {\bf K}_5, \quad \left[
{\bf K}_3, {\bf K}_5 \right] =  - {\bf K}_4, \quad \left[ {\bf
K}_4, {\bf K}_5 \right] = 2 \omega {\bf K}_1 . \nonumber
\end{eqnarray}

{\it Class III} : $m^2 \equiv - \mu^2 < 0, \omega \neq 0$. In this
Class, the KVs are
\begin{eqnarray}
& & {\bf K}_1 = \partial_t, \quad {\bf K}_2 = \partial_z, \quad
{\bf K}_3 = \frac{2 \omega}{\mu} \partial_t + \mu \,
\partial_{\phi}, \nonumber \\& & {\bf K}_4 = - \frac{H}{D} \sin\phi
\, \partial_t + \cos\phi \partial_r - \frac{D'}{D} \sin\phi
\partial_{\phi}, \\ & & {\bf K}_5 = - \frac{H}{D} \cos\phi
\partial_t - \sin\phi \partial_r - \frac{D'}{D} \cos\phi \partial_{\phi} \nonumber
\end{eqnarray}
where the Lie algebra is given by
\begin{eqnarray}
& & \left[ {\bf K}_3, {\bf K}_4 \right] = \mu {\bf K}_5, \quad
\left[ {\bf K}_3, {\bf K}_5 \right] = - \mu {\bf K}_4, \quad
\left[ {\bf K}_4, {\bf K}_5 \right] = \mu {\bf K}_3 . \nonumber
\end{eqnarray}

{\it Class IV} : $m^2 \neq 0, \omega = 0$. In this class, $H = 0$
and $D(r) = \frac{1}{m} sinh(mr)$ for $m^2 > 0$, or $D(r) =
\frac{1}{\mu} sin(\mu r)$ for $m^2 \equiv - \mu^2 < 0$.

\begin{eqnarray}
& & {\bf K}_1 = \partial_t, \quad {\bf K}_2 = \partial_z, \quad
{\bf K}_3 = z \, \partial_t + t \partial_z, \nonumber
\\ & & {\bf K}_4 = \cos\phi \partial_r - \frac{D'}{D} \sin\phi \partial_{\phi}, \\
& & {\bf K}_5 = - \sin\phi \partial_r - \frac{D'}{D} \cos\phi
\partial_{\phi}, \quad {\bf K}_6 = \partial_{\phi} \nonumber
\end{eqnarray}
with the nonvanishing commutators
\begin{eqnarray}
& & \left[ {\bf K}_1, {\bf K}_3 \right] = {\bf K}_2, \quad \left[
{\bf K}_2, {\bf K}_3 \right] = {\bf K}_1 , \nonumber \\& & \left[
{\bf K}_4, {\bf K}_5 \right] = - m^2 {\bf K}_6 , \quad \left[{\bf
K}_5, {\bf K}_6 \right] = {\bf K}_4 , \quad \left[ {\bf K}_6, {\bf
K}_4 \right] = {\bf K}_5 . \nonumber
\end{eqnarray}
It is noticed that when $m = \omega = 0$ the line element
(\ref{metric}) is clearly Minkowskian. It admits the Poincar\'{e}
group.

\end{document}